\documentclass[aps,pra,twocolumn,floatfix]{revtex4-1}
\usepackage{ulem}
\usepackage{amsfonts}
\usepackage{amsmath}
\usepackage{amssymb}
\usepackage{graphicx}
\usepackage{float}
\usepackage[toc,page,header]{appendix}

\begin{document}

\title{Resonance-enhanced collective effect in a triangle arrangement of Rydberg atoms with anisotropic interactions}
\author{Jing Qian}
\affiliation{Department of Physics, School of Physics and Material Science, East China Normal University, Shanghai 200062, P.R. China}

\begin{abstract}
We investigate the collective excitation effect in a scheme where three identical Rydberg atoms are arranged in an equilateral triangular lattice. By using a static electric field polarizing the atomic dipoles, the dipole-dipole interactions between two Rydberg atoms are essentially anisotropic and can even disappear in the several special resonance cases.
For that fact, we observe collectively enhanced excitation probability of single Rydberg atom in resonant areas in the case of strong blockade, and that of double or triple Rydberg atoms in the case of partial blockade. To give more evidences for this collective excitation enhancement, we study the two-body quantum correlation between three Rydberg atoms, as well as the dependence of the blockade radius on the length of triangle sides, which present a good agreement with the excitation properties.
\end{abstract}

\maketitle
\preprint{}

\section{Introduction}

Collective coherent quantum effects existing in a many-body system can be enhanced via strong couplings. 
A typical example is the scheme of Dicke superradiance, in which the excitation of $N$ two-level atoms with a common light obtains an enhanced emission because of coherence \cite{Dick54}. 
And in cavity QED, the cavity-assisted atom-field coupling presents a $\sqrt{N}$ collective enhancement for atomic Bose-Einstein condensates \cite{Colombe07,Brennecke07}. In particular, ensembles of highly excited Rydberg atoms offer a robust platform for studying quantum correlation and collective effect by engineering the strength, range, direction of Rydberg interactions. In Rydberg atoms the competition between laser pumping and interatomic interaction ensures that only one Rydberg atomic excitation can be accommodated within the blockade sphere due to the energy level shifts induced by the surrounding atoms \cite{Comparat10}, giving rise to a $\sqrt{N}$ enhancement to the atom-light coupling \cite{Heidemann07,Dudinnp12,Carmele14}.

Interactions between two Rydberg states can be isotropic van der Waals (vdWs)-type or anisotropic dipole-dipole-type, depending on the magnitude of the applied electric field \cite{Reinhard08}. The vdWs interaction often occurs at large detunings where the dipole moment of atoms is temporary and weak, and it has been directly measured by experiments \cite{Beguin13,Thaicharoen15}. On the other hand, the dipole-dipole interactions (DDIs) in the presence of strong external fields is essentially long-range and anisotropic. Since the angular dependence (anisotropy) of DDIs has been studied in various geometries \cite{Carroll04,Ravets15,Bigelow15}, controlling over the strength of DDIs by varying the magnitude and direction of electric field has found applications in the implementation of quantum information protocols \cite{Saffman10,Saffman16} or in the quantum simulation of  many-body spin systems \cite{Weimer10}.

To focus on the anisotropy of interactions, a two-dimensional (2D) triangular configuration with three atoms is priorly preferred because in the case of $N=3$ some atom pairs necessarily have an interatomic axis that is not aligned along the quantization axis \cite{Kiffner13,Barredo14}. In addition, the DDIs among three interacting atoms can cause breakup or reduction to the blockade effect owing to the cooperative behavior of different energy exchange channels  \cite{Pohl09,Qian13p2}, which further make the excitation properties of three atoms qualitatively differ from that of two-atom systems \cite{Gaetan09,Urban09}. In the present work, we study the angular-dependent Rydberg excitations in a three-atom system in which the strength of DDIs between atoms can be manipulated by the direction of an external electric field. For special directions where the interaction vanishes between two of three atoms, which is so-called {\it resonance} in our paper, we observe enhanced  probability of single collective excitations in the case of full blockade due to the cooperativity. For comparison we also consider the case of partial blockade where we find the double and triple collective state populations are enhanced due to weaker interactions and imperfect blockade effect.

Also, to investigate collective effect between pairs of atoms, we compute the two-atom quantum correlation coefficient and find that the collective excitation enhancement at resonance could significantly affect the correlation properties between atoms, giving rise to positive correlation even in full blockade. Besides, we study the change of Rydberg blockade sphere characterized by the blockade radius, and a strong reduction to it at resonance is found, accounting for the imperfect blockade. Our results give a visualized picture of a deformed three-atom blockade sphere by the effect of angular dependent interactions.

\section{The three-atom model}  \label{Mmo}

\begin{figure}[hbt]
\includegraphics[width=3.2 in]{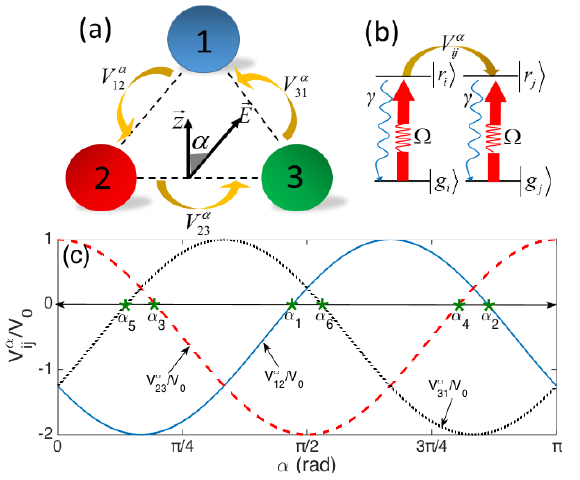}
\caption{(Color online) (a) Schematic representation of three interacting atoms trapped in an equilateral triangular lattice with angular-dependent two-body interactions $V_{12}^{\alpha}$, $V_{23}^{\alpha}$, $V_{31}^{\alpha}$. $\alpha$ is the angle between the quantization axis $\vec{z}$ and the electric field $\vec{E}$. $\vec{z}$ is vertical to the internuclear distance of atoms 2 and 3. (b) Level structure of atoms with ground state $\left\vert g_j \right\rangle$ and highly-excited Rydberg state $\left\vert r_j \right\rangle$ coupled by the effective Rabi frequency $\Omega$. $\gamma$ stands for the spontaneous decay rate in $\left\vert r_j \right\rangle\to\left\vert g_j \right\rangle$ transition. (c) The ratio of resonant dipole-dipole interaction $V_{ij}^{\alpha}$ to the off-resonant interaction $V_0$ as $\alpha\in(0,\pi)$, which are $V_{12}^{\alpha}/V_{0}$ (blue solid), $V_{23}^{\alpha}/V_{0}$ (red dashed), $V_{31}^{\alpha}/V_{0}$ (black dotted). Resonances with the strength of dipole-dipole interactions $V_{ij}^{\alpha}$ between one pair of atoms $i$ and $j$ detuned to be zero are marked by green stars. By the equilateral triangular geometry, from left to right, $\alpha_5=\alpha_m-\pi/6$, $\alpha_3=\pi/2-\alpha_m$, $\alpha_1=\pi/6+\alpha_m$, $\alpha_6=5\pi/6-\alpha_m$, $\alpha_4=\pi/2+\alpha_m$, $\alpha_2=7\pi/6-\alpha_m$, where the "magic angle" is $\alpha_m=\arccos(1/\sqrt{3})$.}
\label{model}
\end{figure}

The geometry of a three-atom model is shown in Fig. \ref{model}(a), in which we consider three identical ultracold Rydberg atoms with the interatomic two-body interactions trapped in a 2D equilateral triangular lattice. In the frozen limit, the center of mass motion of atoms can be negligible with a deep trap depth.
Each atom [see Fig. \ref{model}(b)] modeled by a two-level structure is resonantly excited from the ground state $\left \vert g_j \right \rangle$ to the highly-excited Rydberg state $\left \vert r_j \right \rangle$ by the effective Rabi frequency $\Omega$, decaying spontaneously with rate $\gamma$. This two-level simplification is quite feasible in the system of three-level atoms where the detuning to the intermediate state (not shown) is very large.

In the absence of external fields, two atoms prepared in the same state generally interact via vdWs interactions.
The huge polarizability of Rydberg atoms makes them very sensitive to electric fields, so we apply an electric field $\vec{E}$ to control the atomic transition dipoles. If all dipoles are aligned along the same orientation $\vec{E}$  deviating from the quantization axis $\vec{z}$ by $\alpha$, interactions between two Rydberg states $\left\vert r_i\right\rangle$ and $\left\vert r_j\right\rangle$ are given by the electric dipole-dipole type \cite{Comparat10}, 
\begin{equation}
V_{ij}^{\alpha} =V_0(1-3\cos^2\theta_{ij}^{\alpha}),
\label{Vs}
\end{equation}
with $V_0$ standing for the strength of interaction at off-resonance and $\theta_{ij}^{\alpha}$ the angle between interatomic distance and the electric field, which are $\theta_{12}^{\alpha}=\alpha-\frac{\pi}{6}$, $\theta_{23}^{\alpha}=\frac{\pi}{2}-\alpha$ and $\theta_{31}^{\alpha}=\frac{5\pi}{6}-\alpha$. 

It is worth stressing that this classical DDIs corresponding to two interacting dipole moments \cite{Lahaye09,Comparat10,Zhelyazkova15} fully differs from the resonant DDIs that arise from the degeneracy of Rydberg levels at F\"{o}rster resonance \cite{Vogt06,Ditzhuijzen08,Ryabtsev10,Altiere11,Nipper12}. In the latter case, the resonant interactions between Rydberg atoms emerge through exchange excitation due to nearby degenerate Rydberg levels, and the complex angular dependence of which has been classified and displayed clearly in two individual atoms \cite{Ravets15}. The  goal of our current work focus on the collective excitation properties of three atoms influenced by the angular-dependent DDIs, so it is sufficient to consider simple two-level systems without the inclusion of nearby Rydberg energy levels.

Figure \ref{model}(c) represents the variation of $V_{ij}^{\alpha}/V_0$ as a function of $\alpha$ where $V_{12}^{\alpha}/V_0$, $V_{23}^{\alpha}/V_0$, $V_{31}^{\alpha}/V_0$ are respectively displayed by blue solid, red dashed and black dotted curves.
The concept {\it resonance} is newly introduced in the scheme for defining the location where the DDI between one pair of atoms vanishes, as denoted by $\alpha_k$(k=1,2,...,6) (green stars) in the picture. Accordingly, we assume $V_{12}^{\alpha}=0$ at $\alpha=\alpha_1$ or $\alpha_2$; $V_{23}^{\alpha}=0$ at $\alpha=\alpha_3$ or $\alpha_4$; $V_{31}^{\alpha}=0$ at $\alpha=\alpha_5$ or $\alpha_6$.

With the rotating-wave approximation, the total Hamiltonian $\mathcal{\hat{H}}$ of the system can be written as the sum of single atom Hamiltonian as well as the interactions among them, which is given by ($\hbar=1$)
\begin{equation}	\hat{\mathcal{H}}=\Omega\sum_{j=1}^3(\hat{\sigma}_{gr}^{(j)}+\hat{\sigma}_{rg}^{(j)})+\sum_{ij=12,23,31}V_{ij}^{\alpha}\hat{\sigma}_{rr}^{(i)}\hat{\sigma}_{rr}^{(j)},
	\label{Ham}
\end{equation}
where $\hat{\sigma}_{\alpha \beta}^{(j)}=\left\vert \alpha_j\right\rangle\left\langle \beta_j\right\vert$ with $\alpha,\beta \in (g,r)$ are the operators of atom $j$. 
Incoherent effects such as the spontaneous emission from Rydberg state $\left\vert r_j\right\rangle$ would be included in the Lindblad operator part of the master equation: $\partial_t\hat{\rho}(t)=-i[\hat{\mathcal{H}},\hat{\rho}(t)]+\hat{\mathcal{L}}[\hat{\rho}]$. The Lindblad operator $\hat{\mathcal{L}}[\hat{\rho}]$ is
\begin{equation}
	\hat{\mathcal{L}}[\hat{\rho}]=\frac{\gamma}{2}\sum_{j=1}^3(2\hat{\sigma}_{gr}^{(j)}\hat{\rho}\hat{\sigma}_{rg}^{(j)}-\hat{\sigma}_{rr}^{(j)}\hat{\rho}-\hat{\rho}\hat{\sigma}_{rr}^{(j)}).
	\label{Lind}
\end{equation}
with $\hat{\rho}$ a $2^3\times2^3$ density matrix and $\hat{\mathcal{L}}[\hat{\rho}]$ a sum of independent single atom dissipators. Note that we have assumed a complete set of base vectors composed of $2^3$ three-atom collective state $\left\vert ijk\right\rangle = \{\left\vert ggg\right\rangle$, $\left\vert ggr\right\rangle$, $\left\vert grg\right\rangle$, $\left\vert grr\right\rangle$, $\left\vert rgg\right\rangle$, $\left\vert rgr\right\rangle$, $\left\vert rrg\right\rangle$, $\left\vert rrr\right\rangle\}$ and $P_{ijk}$ is the steady state population of $\left\vert ijk\right\rangle$.

Furthermore, to qualitatively understand the collective excitation properties, in the following numerical explorations, we consider two regimes: full blockade and partial blockade, characterized by the relative strengths between $V_0$ and $\Omega$. If $V_{0}\gg \Omega$ it is full blockade with the single collective Rydberg excitation shared by all atoms, leading to a so-called "superatom" \cite{Weber15,Zeiher15,Honer11}, where the multiple Rydberg excitations are strongly blocked. 
If $V_{0}$ is comparable to $\Omega$ it is partial blockade where the double and triple collective state probabilities obtain a great enhancement \cite{Beguin13,Barredo14}. If  $V_{0}\ll \Omega$ all atoms behave independently and no blockade exists. Since both the sign and strength of real interatomic interaction $V_{ij}^{\alpha}$ are changeable, our proposal can serve as a good candidate for studying both cases.

\section{Excitation enhancement at Resonance}  \label{Prob}

We investigate the excitation probabilities in collective Rydberg states when the dynamics of atoms comes to steady state under the effect of laser driving and spontaneous emission. We use $P_{r_j}$ for describing the steady state probability of atom $j$ to Rydberg state, which are defined by
\begin{eqnarray}
	P_{r_1} &\equiv &P_{rgg}+P_{rgr}+P_{rrg}+P_{rrr},  \\
	P_{r_2} &\equiv &P_{grg}+P_{grr}+P_{rrg}+P_{rrr},  \\
	P_{r_3} &\equiv &P_{ggr}+P_{grr}+P_{rgr}+P_{rrr},
\label{excitpr}
\end{eqnarray}
as well as $P_{jr}$ the collective excitation probabilities of $j$ atoms ($j$ is the atomic number), which are
\begin{eqnarray}
      P_{1r} &\equiv &P_{rgg}+P_{grg}+P_{ggr},  \\
	P_{2r} &\equiv &P_{rrg}+P_{rgr}+P_{grr},  \\
	P_{3r} &\equiv &P_{rrr}.
	\label{excitpro}
\end{eqnarray}

In the simulation we consider the full blockade case with $V_{0}=50$ and the partial blockade case with $V_{0}=5.0$, as well as  $\Omega$ changes within $[0,10]$. All parameters scale with the decay rate $\gamma$. If $\gamma=1.0$MHz and the dispersion coefficient $C_3$ for the DDI is $C_3=2.39 $GHz $\mu$m$^3$ in $^{87}$Rb \cite{Ravets14}, the required interatomic distance (i.e. the side of triangular lattice) is 3.63$\mu$m (full bockade) and 7.82$\mu$m (partial blockade), respectively. Changing the distance between two atoms is feasible in experiment, for example, by controlling the incidence angle of laser beams \cite{Beguin13}.

\subsection{Full blockade}

 \begin{figure}[ptb]
\includegraphics[width=3.5in,height=2.3in]{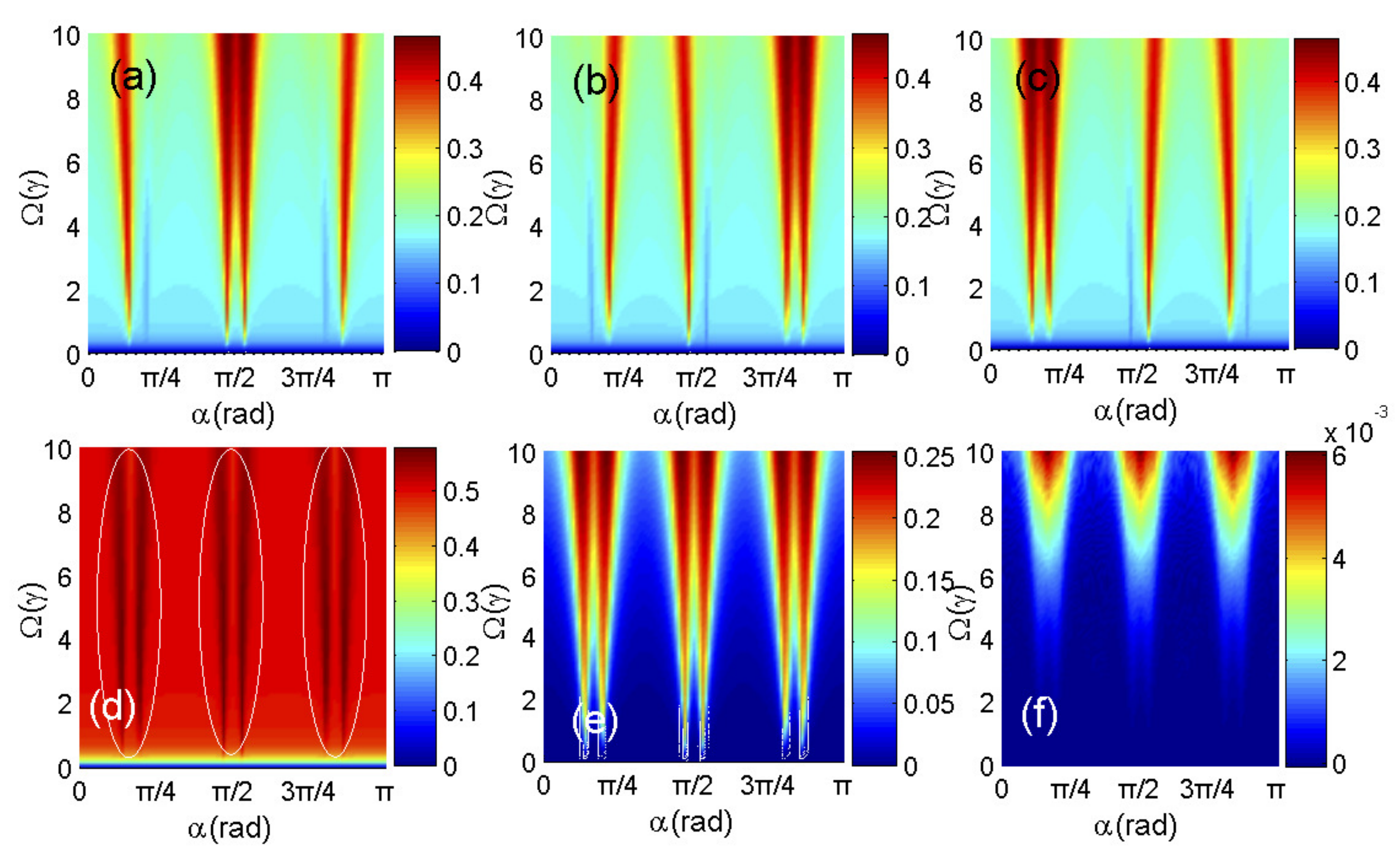}
\caption{(Color online) Steady state population of Rydberg states in the parameter space $(\alpha,\Omega)$ in the full blockade regime. (a)-(f) respectively show $P_{r_1}$, $P_{r_2}$, $P_{r_3}$, $P_{1r}$, $P_{2r}$ and $P_{3r}$.} 
\label{ddint4d}
\end{figure}

To study the steady state and excitation properties in the full blockade with $V_0\gg\Omega$, the solutions from numerically solving the master equation are displayed in Fig. \ref{ddint4d} in which the steady state probabilities $P_{r_j}$ [(a)-(c)] and $P_{jr}$ [(d)-(f)] ($j=1,2,3$ accordingly) are shown in space of $(\alpha,\Omega)$. Generally speaking, six well-contrasted resonances are easily observable at $\alpha=\alpha_{1,2,...,6}$ where the enhanced excitation probabilities are observed. The reason behind that enhancement originates from the vanishing of interatomic interaction at resonance, leading to the breakup of full blockade. More detailedly, in (a)-(c) we display the Rydberg excitation probability of atom $j$ which falls down to a very low level if the interaction between atoms $i$ and $k$ becomes to be zero. For instance, if $\alpha=\alpha_1$ or $\alpha_2$, leading to $V_{12}=0 \ll |V_{23}|, |V_{31}|$, the excitation of atom 3 at $\alpha=\alpha_1$ or $\alpha_2$ is largely suppressed, see (c), as both the strong  $|V_{23}|$ and $|V_{31}|$ working on it. Atoms 1 and 2 have a dominant excitation because only $V_{31(23)}$ works on atom 1(2) and $V_{12}=0$. 
Also $P_{r_j}$ saturating to 0.5 at resonance with the increase of $\Omega$ confirms the excitation property of a single two-level atom \cite{Petrosyan13}.

\begin{figure}[ptb]
\includegraphics[width=3.2in,height=2.4in]{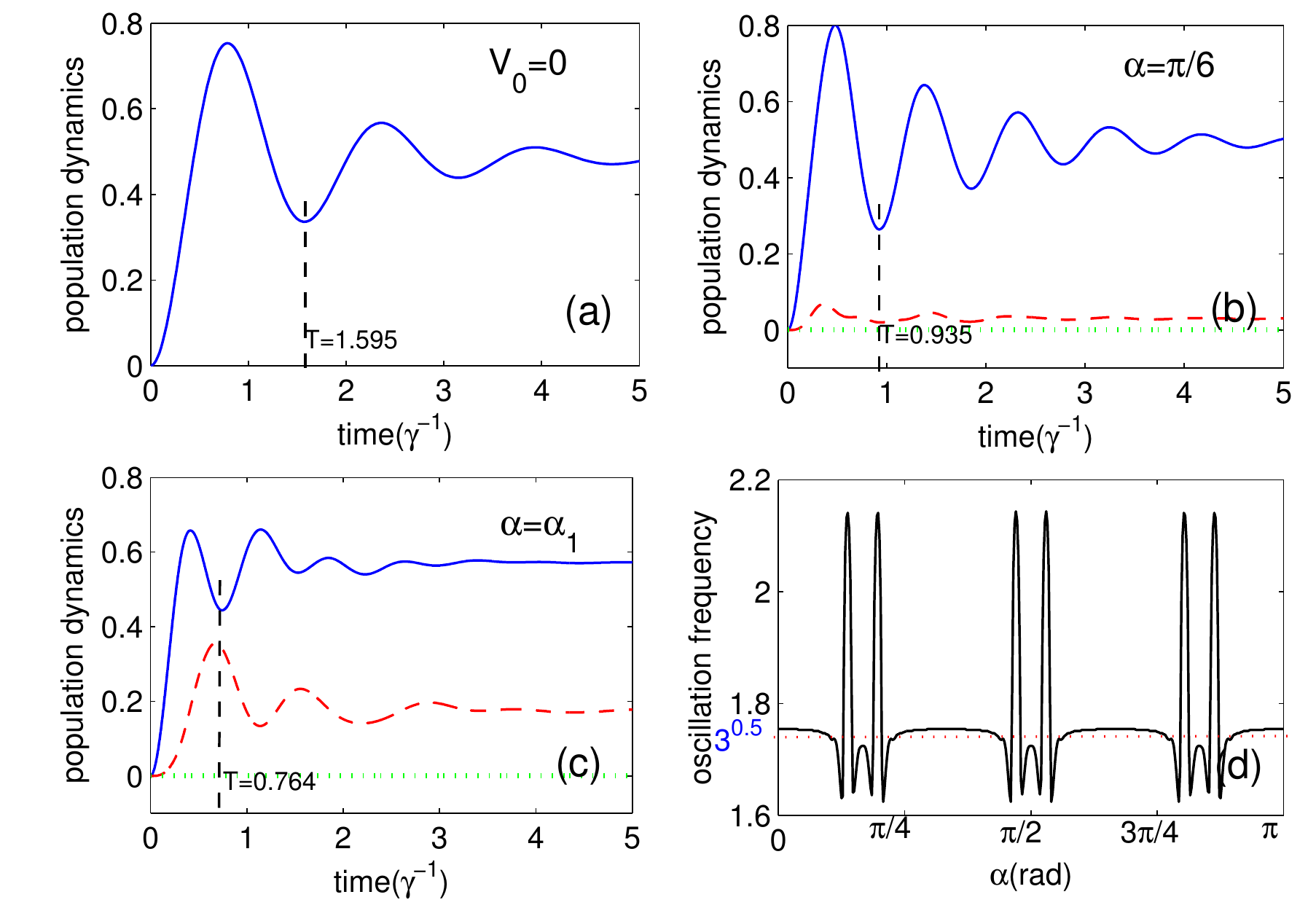}
\caption{(Color online) Realistic population dynamics for three cases: (a) non-interacting single atom with $V_0=0$, (b) the off-resonant case with the collective single (blue solid curve, $P_{1r}$), double (red dashed curve, $P_{2r}$), triple (green dotted curve, $P_{3r}$) Rydberg states, where $\alpha=\pi/6$, (c) is similar to (b) except for $\alpha=\alpha_1$,  presenting the resonant case. (d) The oscillation frequency of single collective Rydberg state with respect to $\alpha$. $\Omega$=2.0, $V_0$=50.0.} 
\label{ddint5d}
\end{figure}

Results in Fig. \ref{ddint4d}(d)-(f) show the excitation probabilities $P_{1r}$, $P_{2r}$ and $P_{3r}$ in collective single, double and triple Rydberg states. To understand more profoundly, for comparison, in Fig. \ref{ddint5d}(b) and (c) we display the realistic population dynamics with $P_{1r}$ (blue solid curve), $P_{2r}$ (red dashed curve) and $P_{3r}$ (green dotted curve) at $\alpha=\pi/6$ (off-resonance) and $\alpha=\alpha_1$ (resonance). Also in Fig. \ref{ddint5d}(a) we show the oscillatory population dynamics of single non-interacting atom. 
At $\alpha=\pi/6$, the resulting $P_{1r}\to 0.5$ and $P_{2r(3r)}\approx 0$ confirm that the full blockade effect with an oscillation frequency $1.71\Omega$ (1.595/0.935$\approx$1.71), close to the expected collective single excitation frequency $\sqrt{3}\Omega$ for three atoms. 
However, turning to the case of resonance, the collective excitation properties are significantly changed by the vanishing of interatomic interaction in one pair of atoms, giving rise to an enhanced excitation probability for double Rydberg state [see Fig. \ref{ddint4d}(e) and Fig. \ref{ddint5d}(c)]. This finding is similar to antiblockade effect where the DDI is compensated by appropriate detunings.
In addition, it is worth noting that in Fig. \ref{ddint4d}(d) and Fig. \ref{ddint5d}(c), the single Rydberg excitation probability $P_{1r}$ is also collectively enhanced at resonance, even transcending the limit 0.5 for single non-interacting atom. Meanwhile, the collective oscillation frequency there increasing to $2.10\Omega$ (1.595/0.764$\approx$2.10) is caused by admixing full blockade and partial antiblockade properties, revealing nontrivial three-atom states. Increasing the number of atoms would improve this effect.

A short representation for the oscillation frequency of single collective excitation $P_{1r}$ is summarized in Fig. \ref{ddint5d}(d) as $\alpha$ changing from 0 to $\pi$. Clearly, the frequency is close to $\sqrt{3}\Omega$ at off resonance places (red dotted curve), but it sharply increases to 2.10$\Omega$ at exact resonances where $\alpha=\alpha_{k}$.

\subsection{Partial blockade}

\begin{figure}[ptb]
\includegraphics[width=3.5in,height=2.3in]{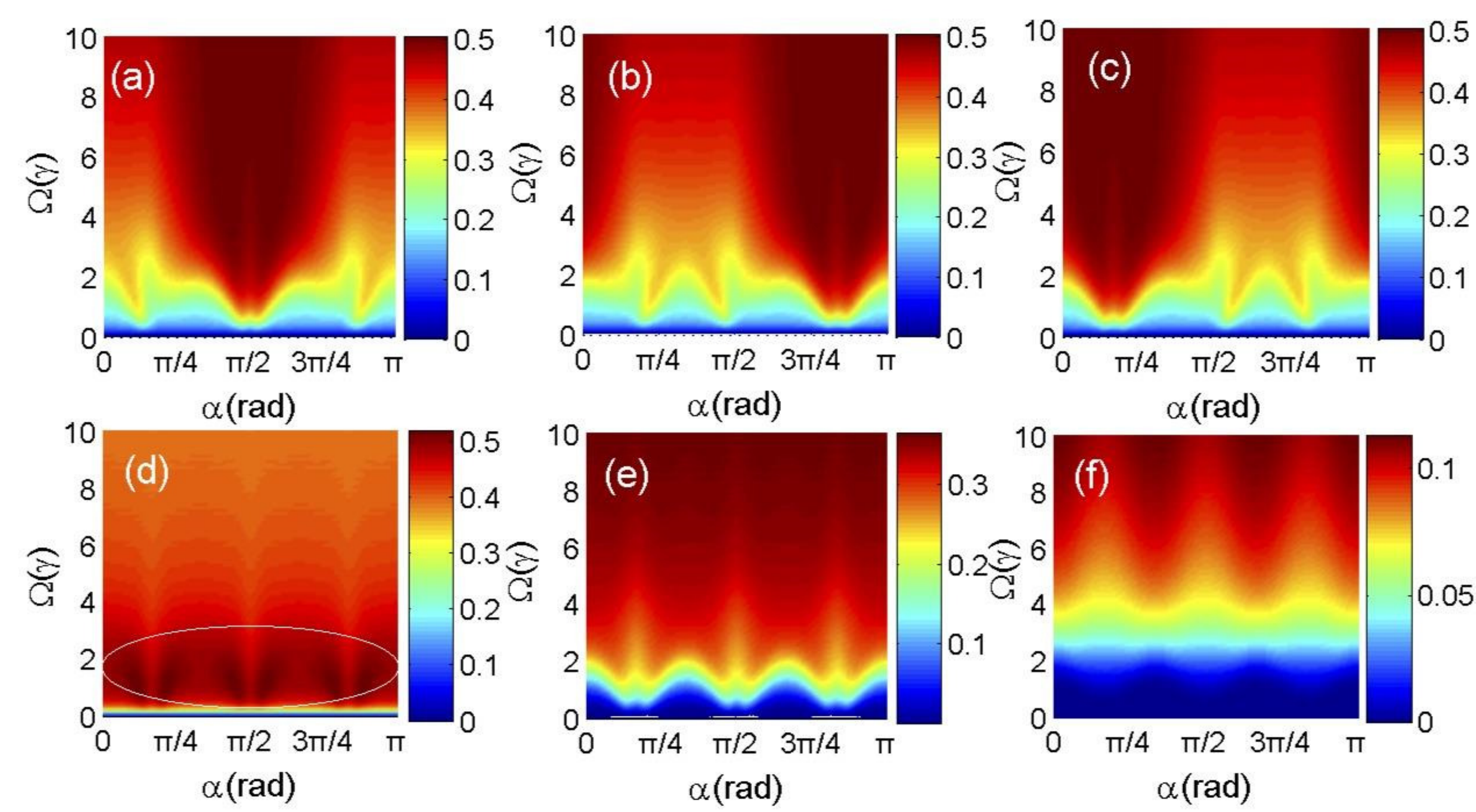}
\caption{(Color online) Similar to figure \ref{ddint4d} except for $V_0=5.0$, presenting a partial blockade case.}
\label{ddint3d}
\end{figure}

\begin{figure}[ptb]
\includegraphics[width=3.2in,height=2.4in]{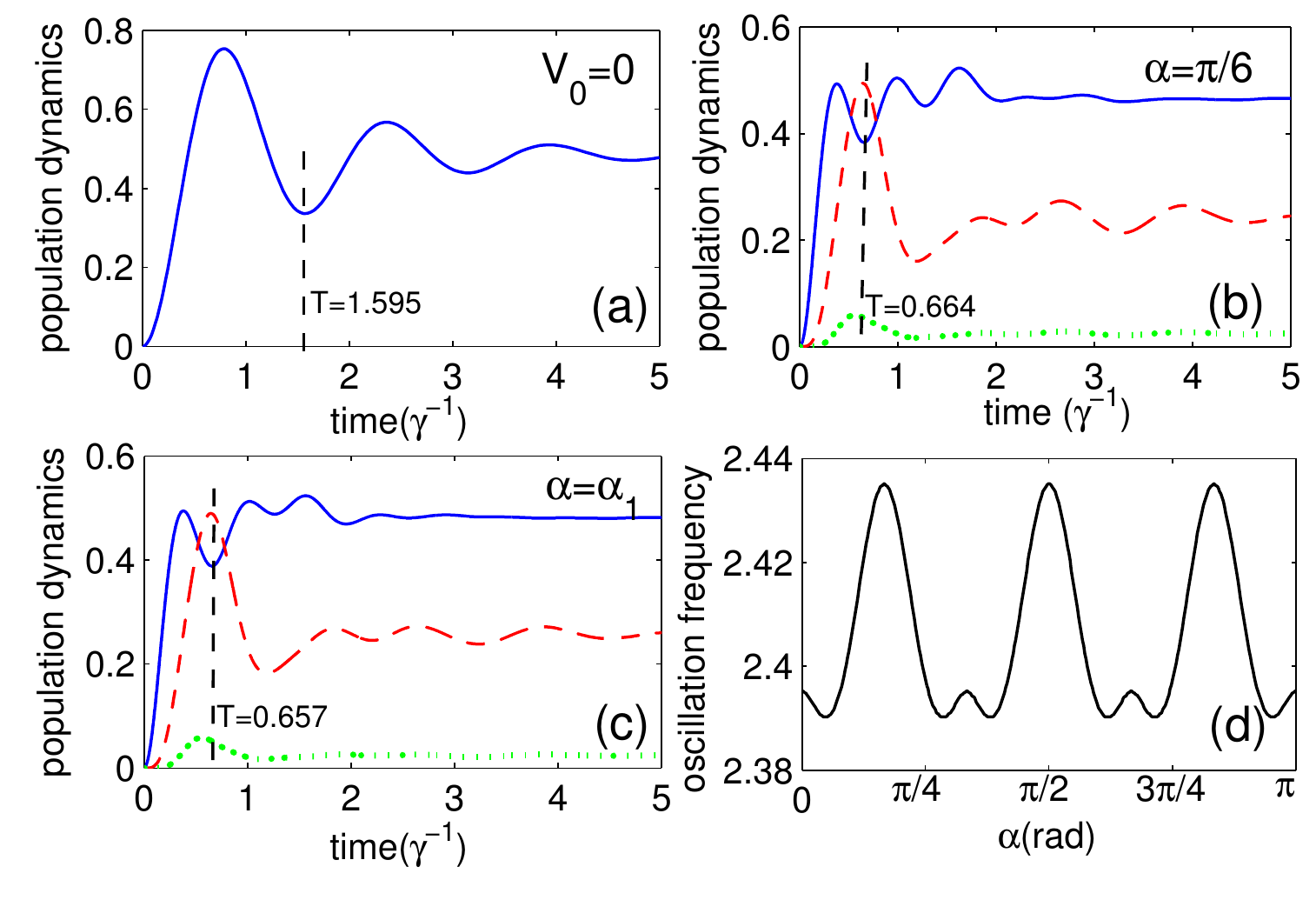}
\caption{(Color online) Similar to figure \ref{ddint5d} except with $V_0$=5.0.} 
\label{ddint6d}
\end{figure}

For comparison, we also study the partial blockade case by choosing $V_0=5.0$, which is comparable to $\Omega$. Differing from the full blockade, we observe resonance can not be well distinguished because the steady state population at off resonance is greatly enhanced. For the excitation of atom $j$, as depicted in Fig. \ref{ddint3d}(a-c), $P_{r_j}$ saturates to 0.5 with $\Omega$ irrespective of $\alpha$. Fig. \ref{ddint3d}(d-f) show that the collective multi-Rydberg excitation probabilities $P_{2r}$ [(e)] and $P_{3r}$ [(f)] increase with $\Omega$, even exceeding the limit given by the excitation of two or three non-interacting atoms, but the single collective excitation rate $P_{1r}$ [(a)] gradually reduces with $\Omega$. To understand this result we also plot the typical population dynamics at $\alpha=\pi/6$ (off resonance, Fig. \ref{ddint6d}(b)) and $\alpha=\alpha_1$ (resonance, Fig. \ref{ddint6d}(c)), where the oscillation frequencies for single collective state both increase to be around 2.40$\Omega$ (quite different from $\sqrt{3}\Omega$). 
In the partial blockade, changing $\alpha$ can not easily affect the collective excitation properties because $V_0$ is comparable to $\Omega$ and plays a less important role. Therefore, Fig. \ref{ddint6d}(b) and (c) represent quite similar population dynamics, and the double and triple Rydberg state probabilities obtain a large enhancement there. In addition, the change of oscillation frequency for single collective excitation with respect to $\alpha$ becomes very slight, see Fig. \ref{ddint6d}(d), again confirming that the cooperation effect preserves unvaried in this case.

\section{Evidence for resonance enhancement}  \label{Evid}

\subsection{Angular-Dependent Quantum Correlation $\delta P_r^{\mu v}$}

Two-body quantum correlation of Rydberg state treated as a good character for describing the collective effect, is typically defined as $\delta\sigma_{\mu v}=\left\langle\sigma_{\mu}\sigma_v\right\rangle-\left\langle\sigma_{\mu}\right\rangle\left\langle\sigma_v\right\rangle$ with $\sigma_{\mu(v)}=\left\vert r\right\rangle\left\langle r\right\vert_{\mu(v)}$. In general, if $\delta\sigma_{\mu v}>0$, it means the double Rydberg state $\left\vert r_{\mu}r_{v}\right\rangle$ trends to be collectively populated; and if $\delta\sigma_{\mu v}<0$, the single Rydberg state $\left\vert r_{\mu(v)}\right\rangle$ prefers to being excited dominatedly \cite{Tony11}. In the former section, we focus on studying the steady state population of Rydberg excitations in three individual atoms; however, the relationship between pairs of atoms with respect to the angular dependence has not been clearly explored. Here, we aim to present the collective effect in double Rydberg excitation at both resonance and off resonance, by considering the two-body quantum correlation coefficient $\delta P_{r}^{\mu v}$ ($\delta P_{r}^{\mu v}=\delta \sigma_{\mu v}$)
\begin{eqnarray}
        \delta P_r^{12}&=&P_{rrg}+P_{rrr}-P_{r_1}P_{r_2}, \\
\label{cora}
        \delta P_r^{23}&=&P_{grr}+P_{rrr}-P_{r_2}P_{r_3}, \\
\label{corb}
        \delta P_r^{31}&=&P_{rgr}+P_{rrr}-P_{r_3}P_{r_1}. 
\label{corc}
\end{eqnarray}

According to definitions of (10)-(\ref{corc}), similarly, if $\delta P_{r}^{\mu v}>0$, the system prefers to populating state $\left\vert r_{\mu}r_{v}\right\rangle$, which possibly occurs at resonance where the vanishing of DDI in one pair of atoms results in a collective excitation for them. On the other hand, if $\delta P_{r}^{\mu v}<0$ it means the system tends to be individually excited to $\left\vert r_{\mu}\right\rangle$ or $\left\vert r_{v}\right\rangle$ because of the interaction-induced full blockade effect.  It is worth stressing that with an angular dependent DDI the system enables an ideal platform for investigating both positive and negative correlations with the help of a merely $\alpha$ adjustment. As for isotropic vdWs type interactions except for $nD$ Rydberg state (the vdWs interaction on $D$-state is anisotropic \cite{Tresp15}) , it is difficult to realize the coexistence of positive and negative correlations in one atom pairs.

\begin{figure}[ptb]
\includegraphics[width=3.5in,height=4.1in]{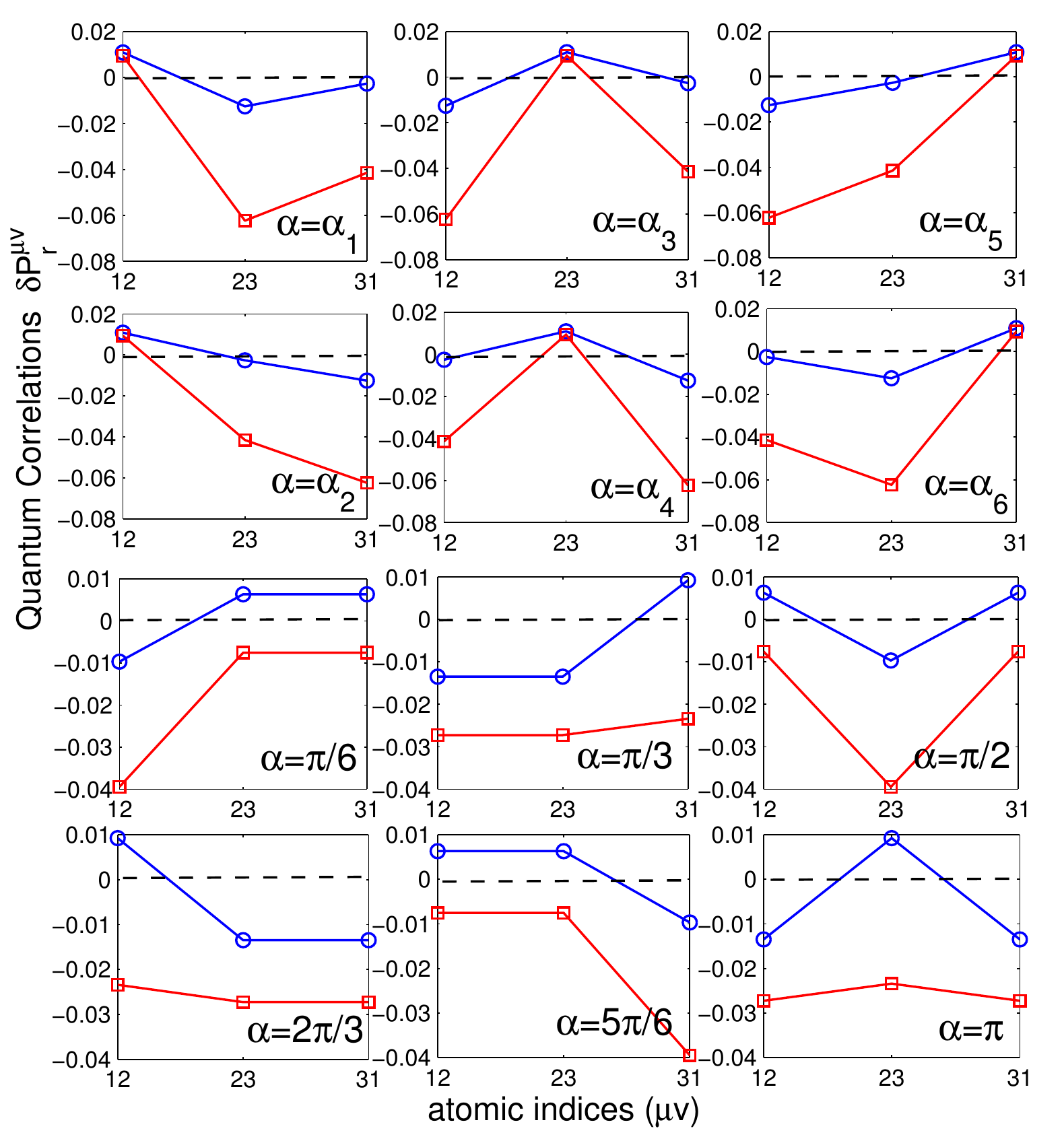}
\caption{(Color online) Quantum correlation coefficient $\delta P_{r}^{\mu v}$ for atom pairs $\mu v=12, 23, 31 $ in the three-atom system. The first and second rows present the resonances with $\alpha=\alpha_{k}$ and the third and fourth rows the off resonances with $\alpha=k\pi/6$ ($k=1,2,...,6$). Results for describing the full blockade with $V_0=50.0$ are denoted red curve with squares, and for the partial blockade with $V_0=5.0$ are denoted blue curve with circles. $\Omega=5.0$.
}
\label{correlation}
\end{figure}

Figure \ref{correlation} represents the two-body correlation coefficient $\delta P_r^{\mu v}$ with atom pairs $\mu v=$12, 23, 31 for given $\alpha$. For comparison, results from full blockade and partial blockade are denoted red curve with squares and blue curve with circles, respectively. Generally speaking, $\delta P_r^{\mu v}$ is more negative (prior individual excitation) in the case of full blockade due to stronger interaction.
Furthermore, in the upper two panels, we find that if $\alpha=\alpha_1$ or $\alpha_2$ (the first column) with $V_{12}^{\alpha_{1(2)}}=0$, $\delta P_r^{12}$ is small positive and $\delta P_r^{23}$ and $\delta P_r^{31}$ are both negative.
This confirms the fact that the vanishing of DDI in one atom pair can lead to a collectively enhanced double Rydberg excitation for that pair, overcoming the two-atom blockade. 
As for a smaller $V_0$, it gives rise to same positive correlation because the interaction between atoms 1 and 2 persists zero at $\alpha=\alpha_1$ or $\alpha_2$. Whereas other negative correlation coefficients for pairs 23 and 31 are deeply growing as $V_{23}^{\alpha}$ and $V_{31}^{\alpha}$ proportional to $V_0$ decrease.
Same results can be obtained when $\alpha=\alpha_{3(4)}$ (the second column) and $\alpha=\alpha_{5(6)}$ (the third column).

Turning to the off resonance where $\alpha=k\pi/6$ ($k=1,2,...,6$) is considered, it is understandable to see that correlation $\delta P_r^{\mu v}$ persists negative values for the full blockade, as denoted red curve with squares in the lower two panels of figure \ref{correlation}. Because the off-resonant DDI (typically comparable to $V_0$) being larger than $\Omega$ would lead to high probability of single-atom excitation, which is confirmed by the suppression of multiple Rydberg excitations in off-resonant full blockade regimes [see Fig. \ref{ddint5d}(b)]. Lowering the strength of $V_0$ will lead to the appearance of positive correlations due to the improving collective effect of multi-Rydberg excitations when entering the partial blockade regime.

\subsection{Angular-Dependent Blockade Sphere Radius $r_b$}

A Rydberg-blockade sphere can provide an intuitive image for explaining the blockade effect, within which a collective single atom excitation shared by the entire atomic ensemble is realized, giving rise to a strong suppression for multi-Rydberg state probability  \cite{Singer04,Dudin12}. The blockade sphere can be well characterized by the blockade radius $r_b$ which is a significant criteria for distinguishing the full blockade and the partial blockade. Note that the mean interatomic distance $r$ describing the triangle side length in our scheme is controllable via realistic experimental tools, so it will be meaningful to study the change of blockade radius with respect to different polarizing angles $\alpha$. Typically, at $r\ll r_b$ it means full blockade and the double Rydberg excitation $P_{2r}\approx 0$; at $r\approx r_b$ it is partial blockade with $P_{2r}$ a finite value; at $r\gg r_b$ every atom shows an independent excitation with $P_{2r}\approx 0.25$. In the numerical simulation we continuously adjust $r$ from 0 to $+\infty$ and find $P_{2r}$ increases to saturation finally, during which the blockade radius $r_b$ is defined by
\begin{equation}
P_{2r}(r_b,\alpha)=\frac{1}{2}[P_{2r}(r=+\infty,\alpha)-P_{2r}(r=0,\alpha)], \\
\label{defRb}
\end{equation}
The double Rydberg excitation probability $P_{2r}$ at $r=r_b$ equals to the half of the subtraction of the maximal-distance ($r=+\infty$) and minimal-distance ($r=0$) probabilities for a given $\alpha$.

\begin{figure}[ptb]
\centering
\includegraphics[width=3.5in,height=1.5in]{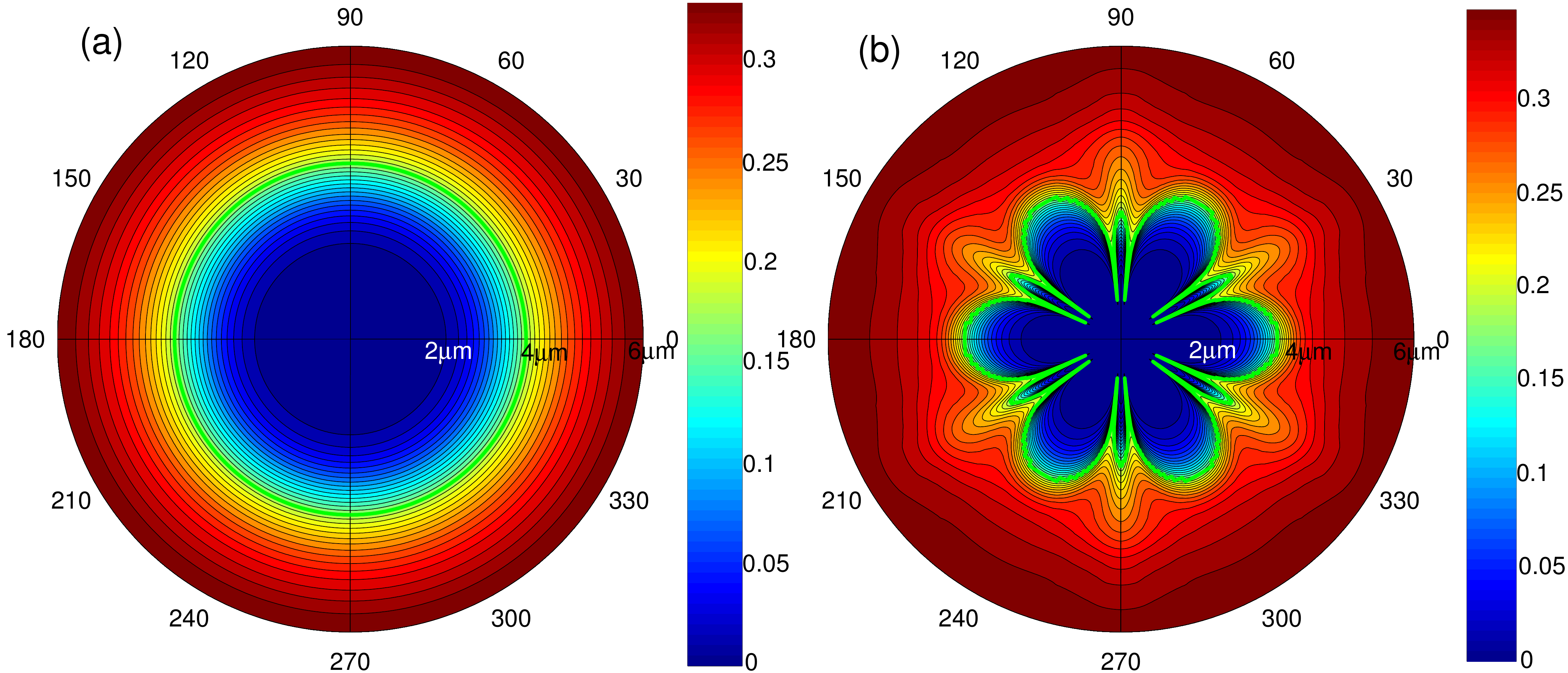}
\caption{(Color online) Representation of the angular dependence of the blockade radius $r_b$ in polar coordinates with respect to $\alpha$, labeled green curves in (b). By symmetry, the points at angles at $\alpha+180$ are taken to be identical to the points at $\alpha$ belonging to $[0,\pi]$. For comparison, a spatially isotropic blockade radius $r_b^{iso}$ with uniform DDI $V_{ij}(r,\alpha)=V_0(r)$ is also shown, labeled green circle in (a). The double excitation probability $P_{2r}$ is displayed by the color-level diagram from 0 (dark blue) to 0.35 (dark red). $\Omega$=5.0. The radial axis is roughly labeled by the interatomic distance $r=(2.0,4.0,6.0)\mu m$ when $\gamma=1.0$MHz is assumed.
}
\label{blockr}
\end{figure}

For capturing a clear physical understanding, in Fig. \ref{blockr}(b), we plot an $\alpha$-dependent three-atom blockade sphere (denoted by green curve) with its radius $r_b$ in the polar coordinate space $(r,\alpha)$, where the probability $P_{2r}$ is shown by a color-level diagram. For comparison, a simple case with the isotropic interaction $V_{ij}^{\alpha}=V_0(r)$ is also computed in Fig. \ref{blockr}(a) where the obtained blockade sphere shows a uniform green circle with radius $r=r_b^{iso}$. In general, $r_b^{iso}$ is unvaried with $\alpha$, but $r_b$ strongly depends on $\alpha$ due to the anisotropy of interatomic interaction. More specifically, in Fig. \ref{blockr}(b) around $\alpha=(2k-1)\pi/6$, we observe that a large reduction to $r_b$, representing the breaking of full blockade by an enhanced excitation of double Rydberg state near resonance (see $P_{2r}$ values there). Beyond resonance, $r_b$ reveals a gradual increase. Especially, we find $r_b\approx r_b^{iso}$ at $\alpha=k\pi/3$, accounting for the maximal DDIs improving the strength of blockade at these off-resonant regimes.

In addition, we find that as the increase of interatomic distance $r$, $P_{2r}$ slowly grows from 0 (dark blue) to more than 0.3 ($>$0.25, dark red), as predicted by a color-level diagram in figures \ref{blockr}(a) and (b), in which the system goes through the full blockade ($P_{2r}\approx 0$), the partial blockade ($0<P_{2r}<0.25$) and the no blockade ($P_{2r}\approx0.25$) regimes. In both isotropic and anisotropic interaction cases, the maximal $P_{2r}$ being larger than 0.30 at $r\to\infty$ is caused by the coherence among three different interaction channels, giving rise to collective excitation, which is absent in the case of two-atom scheme where the maximal $P_{2r}$ is always 0.25 when two atoms do not interact with each other.

\section{Conclusion} \label{Con}

In conclusion, we have studied the anisotropic Rydberg excitation of three two-level atoms trapped in a triangular optical lattice with angular dependent DDIs. Via controlling over the orientation of atomic dipoles by an electric field, we find that the spatially anisotropic interactions could lead to strong resonance at which the single and multiple Rydberg-state excitation probabilities will be coherently enhanced. 
Specifically, different from the isotropic arrangement of Rydberg atoms,
in the full blockade we observe a clear collective single Rydberg excitation enhancement at exact resonance because the vanishing of interaction between one pair of atoms causes partial breakup for the full blockade effect, resulting in imperfect blockade. In the partial blockade the double and triple Rydberg excitation enhancement are also identified due to the strong collective effect.

To verify and understand the enhanced Rydberg excitation, two evidences are presented. One is the two-body correlation between Rydberg atoms, by which we could determine which pair of atoms has a prior excitation by the effect of varied interatomic interactions. The other evidence shows an angular dependent blockade sphere in which a significant reduction to radius at resonance is observed, confirming the enhanced collective Rydberg effect there. In the future we could consider the anisotropic DDIs and collective excitation effect with more than three atoms or in different 2D geometries \cite{Nogrette14}.

This work was supported by the National
Natural Science Foundation of China
under Grant No. 11474094, No. 11104076, and the Specialized Research Fund for the Doctoral Program of Higher Education Grant No. 20110076120004.

\end{document}